# INVARIANCE OF THE LOCAL SEGMENTAL RELAXATION DISPERSION IN POLYCYCLOHEXYLMETHACRYLATE / POLY-α-METHYLSTYRENE BLENDS

C.M. Roland*, R. Casalini*,†

*Naval Research Laboratory, Chemistry Division, Code 6120, Washington DC  20375-5342
†George Mason University, Chemistry Department, Fairfax VA  22030

**Abstract**

Dielectric spectroscopy was carried out on polycyclohexylmethacrylate (PCHMA) and its blend with poly-α-methylstyrene (PaMS) as a function of temperature and pressure. When measured at conditions whereby the local segmental relaxation time for the PCHMA was constant, the dispersion in the loss spectra had a fixed shape; that is, the relaxation time determines the breadth of the relaxation time distribution, independently of $T$ and $P$. This result is known for neat materials and could be observed for the blend herein due to the nonpolar character of the PaMS and the degree of thermodynamic miscibility of the blend.

## 1. Introduction

The most striking aspect of the dynamic properties of glass-forming materials and the immediate cause of their vitrification is the divergence of the relaxation time, $\tau_\alpha$ (structural or, for polymers, local segmental relaxation time), with decreasing temperature or increasing pressure. Theories of the glass transition, variously based on configurational entropy [1,2,3,4] or free volume [5,6,7], attempt to account for the variation of $\tau_\alpha$ with thermodynamic conditions. There are other properties that exhibit interesting behavior in the supercooled regime, but these are not usually addressed by theories. In particular, the dispersion of the relaxation times is ignored, even though it is strongly correlated with $\tau_\alpha$. In fact, the latter uniquely defines the breadth of the dispersion, so that for a given value of $\tau_\alpha$ the shape of the α-dispersion is constant [8,9,10,11].

When two polymers are mixed, the dispersion broadens due to concentration fluctuations [12]. This effect is much greater than for mixtures of small molecules, due to the more limited miscibility (smaller mixing entropy) of polymers. If the local segmental motions of the blend components are sufficiently different, concentration fluctuations can give rise to dynamic heterogeneity, wherein each component exhibits a distinct relaxation peak [13,14]. The question than arises as to whether the relationship between $\tau_\alpha$ and the dispersion [8,9,10,11] is maintained, which can only be addressed if the dispersion is due to just one component of the blend. This requires that dynamic heterogeneity causes the component dynamics to be well separated or, the more usual case, that only one component contributes to the measured response due to large difference in the respective susceptibilities. In dielectric spectroscopy this is realized in blends of a polar and a nonpolar polymer.

In this work we describe dielectric relaxation measurements at ambient and elevated pressure on polycyclohexylmethacrylate (PCHMA) with poly-α-methylstyrene (PaMS). PCHMA and PaMS form miscible blends, which have lower critical solution temperatures exceeding 560 K (approaching decomposition temperatures) even for high molecular weight components [15]. Since the PaMS is relatively nonpolar, the dielectric response of the blend is dominated by the PCHMA. This allows assessment of the blend's conformance to the general result for neat glass-forming materials that the α-dispersion is invariant to $T$ and $P$ for constant $\tau_\alpha$.

## 2. Experimental

The PCHMA (from the Polymer Source) and PaMS (from Polymer Standards Service USA) had respective weight average molecular weights of 3.4 kg/mol and 1.5 kg/mol, with polydispersities equal to 1.14 and 1.29; both were used as received. A 50% by weight blend was prepared by dissolution in chloroform, followed by drying in vacuo one week at RT.

Dielectric measurements were carried out with a parallel plate geometry (2 cm diameter and 50 micron Teflon spacers), with the sample molded between the electrodes at ~430 K and light pressure. Spectra were obtained using an IMASS time domain dielectric analyzer ($10^{-3}$ to $10^3$ Hz) and a Novocontrol Alpha Analyzer ($10^{-2}$ to $10^6$ Hz). For measurements at elevated pressure, the sample capacitor assembly was contained in a Manganin cell (Harwood Engineering), with pressure applied using a hydraulic pump (Enerpac) in combination with a pressure intensifier (Harwood Engineering). Pressures were measured with a Sensotec tensometric transducer (resolution = 150 kPa). The sample assembly was contained in a Tenney Jr. temperature chamber, with control to within ± 0.1 K at the sample.

## 3. Results
### *3.1 Neat Components*

Figure 1 displays the α-peak for neat PCHMA for various temperatures at atmospheric pressure, along with fits to the transform of the Kohlrausch-William-Watts function [16]

$$\phi(t) = \exp[-(t/\tau_\alpha)^\beta] \qquad (1)$$

The function adequately describes the peak over the central portion. Deviations can occur at higher frequencies due to secondary processes [17,18,19] and at lower frequencies due to d.c. conductivity, $\sigma$. In Figure 2 the best-fit values of the stretch exponent $\beta$ are plotted *vs.* the frequency of the maximum of the *α*-dispersion. There is a systematic peak broadening with decreasing frequency (lower temperature or larger $\tau_\alpha$). This sensitivity to temperature becomes more significant closer to $T_g$ (last datum at $\tau_\alpha \sim 3$ s). In general, $\beta$ decreases slowly with decreasing temperature past the dynamic crossover, after which the change is more rapid [20]. The data in Figs. 1 and 2 demonstrate that the α-peaks for different values of $\tau_\alpha$ do not superpose. Similar results are obtained for neat PaMS (not shown): At ambient pressure over the range from $T$ = 336 to 351K, the stretch exponent in eq.(1) increases from 0.30 to 0.32; that is, time-temperature superpositioning also fails for PaMS.

Figure 3 shows α-dispersion measured at various conditions of *T* and *P*. The latter were chosen to give nearly constant values of $\tau_\alpha$, whereupon the peaks for a given $\tau_\alpha$ superpose. Thus, as seen previously for various glass-forming liquids and polymers, the shape of the α-relaxation function depends only on the relaxation time. Notwithstanding the superposition of the α-peak, the d.c. conductivity response observed toward lower frequencies does not superpose. However, a portion of this arises from an almost pressure independent background conductivity that is likely unrelated to the sample. Generally the conductivity has not been found to be defined by $\tau_\alpha$ in the manner of breadth of the α-dispersion [9,10]. Different dynamics for the α-dispersion and $\sigma$ is a manifestation of the decoupling of translational and orientational dynamics [21,22,23,24,25].

*3.2 PCHMA/PaMS Blend*

Because of the low polarity of the PaMS, the α-peak in the blend reflects only local segmental relaxation of the PCHMA component. The peak is broadened relative to the spectra for neat PCHMA due to concentration fluctuations (distribution of environments with varying local composition); however, the dispersion for any given PCHMA segment relaxing in its local environment should have a constant shape at fixed $\tau_\alpha$. This means that the peak measured for the blend, which is the cumulative response of all PCHMA segments, will be constant at fixed $\tau_\alpha$ provided concentration fluctuations are constant with respect to *T* and *P*. This is expected since for the low molecular weights herein, the PCHMA/PaMS blend is very far from its spinodal [15].

In Figure 4 are displayed the α-dispersions measured for the blend under various conditions of *T* and *P*, each corresponding to one of three different $\tau_\alpha$. For a given value of the latter, the dispersion has a constant shape; that is, the breadth of the segmental relaxation is a function of the relaxation time, consistent with the superpositioning behavior of neat PCHMA (Fig. 3). As shown in the ambient pressure data in Figure 5, the peak for the blend systematically broadens with increasing $\tau_\alpha$; thus, the superpositioning of the α-dispersion in Fig. 4 is non-trivial.

**4. Discussion**

The invariance of the α-peak shape for a blend to *T* and *P* at constant value of the α-relaxation time has been shown previously for a block copolymer of polyisoprene and polyvinylethylene [26] and for polyvinylmethylether in blends with poly-2-chlorostyrene [27] and polystyrene [28] (although the latter result is less significant since there is negligible change in the peak with $\tau_\alpha$, unlike the results herein in Figs. 1 and 2). This superpositioning is in accord with the general behavior of neat glass-forming materials [8,9,10,11]. However, in blends concentration fluctuations (i.e., the distribution of local environments) may change with $T$ and $P$. This would alter the shape of the loss peak, since the local environment of a given segment governs the degree of intermolecular cooperativity of its segmental dynamics and hence the shape of α-relaxation function [29,30,31,32]. However, the thermodynamic stability of the PCHMA/PaMS blend minimizes this effect. Moreover, concentration fluctuations exert a minor effect on the dispersion: the peaks for the blends in Fig. 4 are ≤ 0.5 decades broader (FWHM) than for neat PCHMA (Fig. 3).

The $T$-$P$ superpositioning of the α-dispersion at fixed value of $\tau_\alpha$ for neat materials and blends has significant implications. It means that the relaxation time determines the breadth of the dispersion. This breadth, as characterized for example by the stretch exponent $\beta$, is in turn correlated with many dynamic properties of glass-formers. For example, it is well know that the fragility at fixed pressure, defined as $\left.\frac{d \log \tau_\alpha}{T_g dT^{-1}}\right|_{T_g}$, is inversely correlated with $\beta$ [33], with the correlation being strongest for a given family of materials [34,35,36,37] and only maintained at constant pressure [38]. $\beta$ also correlates with the dependence of $\tau_\alpha$ on the scattering vector [39,40,41,42,43].

Another relationship of the breadth of the dispersion is to the dynamic crossover. The crossover refers to a characteristic temperature $T_B$ at which the temperature dependence of $\tau_\alpha$ changes, as seen in derivative plots [44,45]. When $T_B$ is traversed from above, the rate of change of $\beta(T)$ with temperature increases significantly [46]. It has been found that the dynamic crossover occurs at a fixed value of $\tau_\alpha$; that is, higher pressure increases the crossover temperature but $\tau_\alpha(T_B)$ remains the same [47,48,49]. The superpositioning of the α-relaxation peak (Figs. 3 and 4) means that the shape of the dispersion is also constant at the dynamic crossover, independent of $T$ and $P$.

Notwithstanding its significance to the properties of glass-forming liquids and polymers, the dispersion of the structural relaxation is not addressed by most theories of the glass transition; these restrict their attention primarily to the temperature and pressure dependences of $\tau_\alpha$. An exception to this statement is the coupling model of Ngai [50]. From the singular assumption that the cooperative dynamics that govern structural relaxation arise at some fixed value of time, $t_c$ (~ 2 ps for glass-forming materials), the coupling model derives an expression relating $\tau_\alpha$ and $\beta$

$$\tau_\alpha = (\tau_0 t_c^{\beta-1})^{1/\beta} \tag{2}$$

In this equation $\tau_0$ is the non-cooperative relaxation time of the model, which can be identified with the Johari-Goldstein relaxation time, $\tau_{JG}$ [51,52,53]. Since $t_c$ is constant (as assumed by the coupling model and verified experimentally [54,55,56,57]), eq.(2) implies that $\tau_\alpha$ uniquely defines the exponent $\beta$, provided $\tau_0$ is constant for given $\tau_\alpha$. For many different materials a correlation between $\tau_\alpha$ and $\tau_{JG}$ ($\approx \tau_0$), consistent with eq.(2), has been found [51,52,53]. Data showing that for a given material $\tau_{JG}$ is constant for constant $\tau_\alpha$ are limited, one example is benzoin-isobutylether [58]. In addition the excess wing of several van der Waals liquids superpose at constant $\tau_\alpha$ [8,9,10]. This excess wing reflects a JG relaxation that overlaps the primary structural peak [59,60,61].

The fact that $\tau_{JG}$ ($\approx \tau_0$) is constant at fixed $\tau_\alpha$, together with eq.(2), leads to the prediction that the α-dispersion will be constant, independent of thermodynamic conditions, at fixed $\tau_\alpha$. Application of the coupling model specifically to PCHMA, however, is complicated by the possibility of internal plasticization of the segmental relaxation ("hindered glass transition") due to the bulky alkyl group in this polymer [62,63]. This effect, found in other polyalkylmethacrylates, broadens the dispersion (smaller $\beta$), leading to deviations from eq.(2) [64]. Measurements of the JG-process in PCHMA and its blend with PaMS and their analysis in terms of the coupling model are in progress and will be reported at a later date.

Finally, we note that there is a class of materials for which the α-peak does not superpose at constant $\tau_\alpha$, hydrogen bonded materials. Changes in temperature and pressure can change the degree of H-bonding, which alters the peak shape in a way unaccounted for by fixing the value of $\tau_\alpha$. This breakdown of superpositioning has been shown in neat H-bonded materials, such as glycerol [65] and polypropylene glycol [61], and in blends in which there is H-bonding between components; e.g., polyvinylphenol (PVPh) with poly(vinyl ethyl ether) (PVEE) [66] and PVPh with poly(ethylene-co-vinyl acetate) [67]. Interestingly, in the former case higher pressure broadens the dispersion at constant $\tau_\alpha$, while in the latter pressure causes the α-peak to narrow.

## 5. Summary

For both neat PCHMA and its blend with PaMS, the shape of the dispersion in the dielectric loss due to local segmental relaxation is found to be invariant to thermodynamic conditions when $\tau_\alpha$ is maintained constant. While this is a general result for neat polymers as well as molecular glass-formers, there is limited data showing the phenomenon for blends because of complications due to dynamic heterogeneity, wherein the component dynamics have different *T*- and *P*-dependences, and to concentration fluctuations, which can change with thermodynamic conditions and thereby alter the peak shape. The PCHMA blend measured herein is well removed from its spinodal, so that concentration fluctuations are suppressed, and the dielectric response is dominated by the more polar PCHMA; thus, superpositioning of the α-process can be observed.

**Acknowledgements**

We thank K.L. Ngai for stimulating discussions. The work was supported by the Office of Naval Research.

Figure Captions

Figure 1. Dispersion in the dielectric loss of neat PCHMA due to local segmental relaxation. Temperatures (K) are from left to right: 360.2, 363.1, 369.0, 374.6, 391.0, 398.9, and 407.1. Solid lines are fits to the transform of eq.(1). The

Figure 2. Stretch exponent determined from the fit of eq.(1).to the spectra in Fig. 1. The abscissa corresponds to the frequency of the loss maximum. The corresponding FWHM of the peak varies from less than 2.3 decades to 2.6 decades with increasing $\tau_\alpha$.

Figure 3. Dispersion in dielectric loss for neat PCHMA for conditions under which the peak frequencies were almost equal (spectra were shifted ≤ 20%). The peak heights were also adjusted slightly to superpose the maxima.

Figure 4. Dispersion in dielectric loss for PCHMA/PaMS blend for conditions under which the peak frequencies were almost equal (spectra were shifted ≤ 20%). The peak heights were also adjusted slightly to superpose the maxima.

Figure 5. Breadth of the α-peak for the blend at ambient pressure as a function of the frequency of the peak maximum.

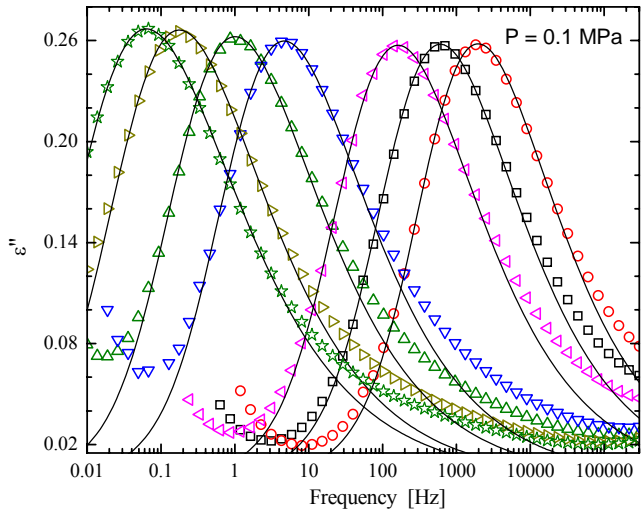

Figure 1.

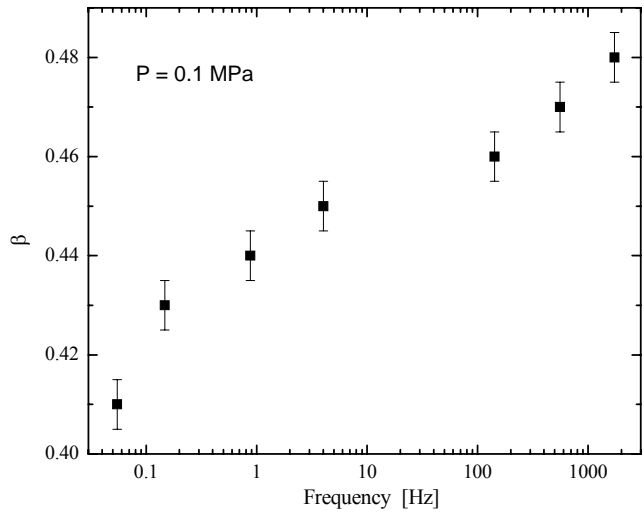

Figure 2

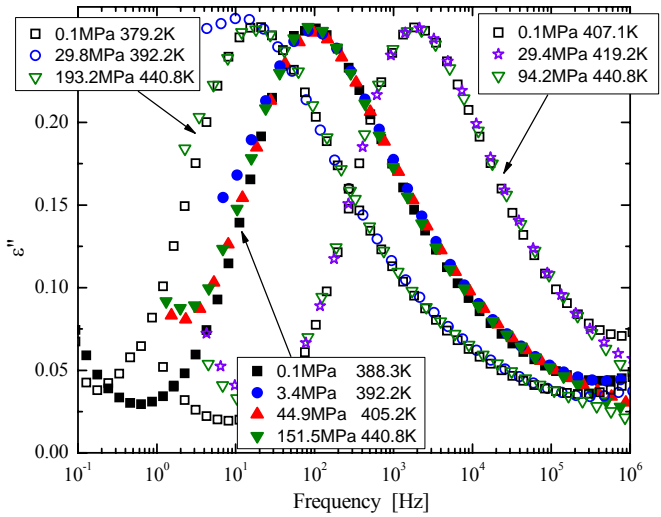

Figure 3

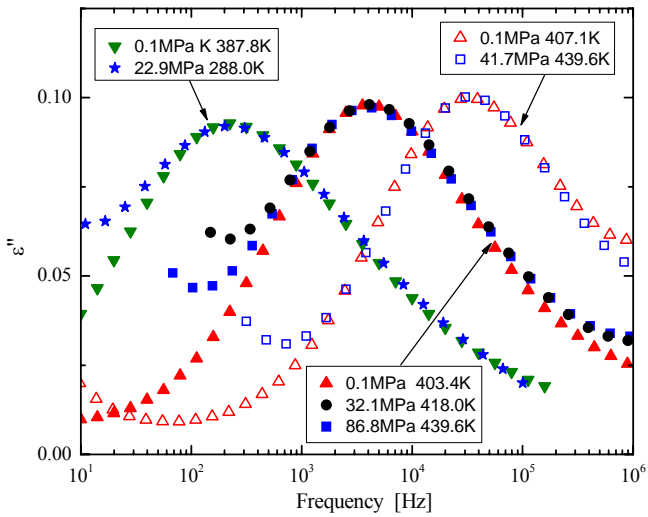

Figure 4

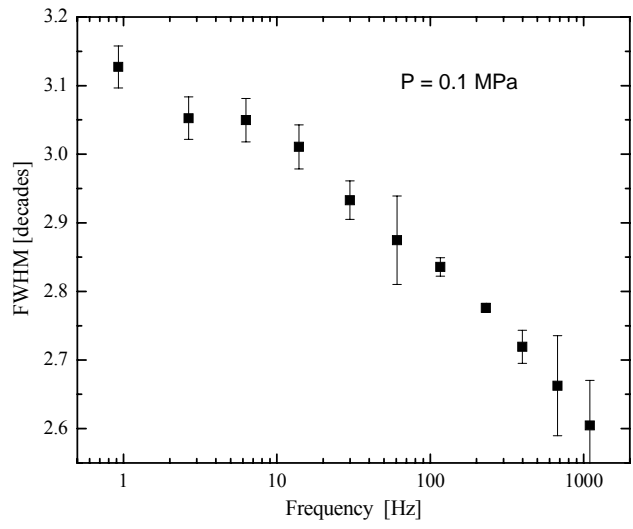

Figure 5